\documentclass[twocolumn]{aastex631}
\usepackage{amsmath}
\usepackage{ulem}

\received{November 16, 2021}
 \accepted{\today}

\submitjournal{RNAAS}

\shorttitle{Two New {\rm roAp} Stars from TESS}
\shortauthors{Jayaraman et al.}
\graphicspath{{./}}
\begin{document}

\title{Two New roAp Stars Discovered with TESS}

\author[0000-0002-7778-3117]{Rahul Jayaraman}
\affiliation{MIT Kavli Institute and Department of Physics, 77 Massachusetts Avenue, Cambridge, MA 02139}

\author[0000-0002-1015-3268]{Donald W. Kurtz}
\affiliation{Centre for Space Research, Physics Department, North West University, Mahikeng 2745, South Africa}
\affiliation{Jeremiah Horrocks Institute, University of Central Lancashire, Preston PR1 2HE, United Kingdom}

\author{Gerald Handler}
\affiliation{Nicolaus Copernicus Astronomical Center of the Polish Academy of Sciences, Bartycka 18, 00-716 Warsaw, Poland}

\author{Saul Rappaport}
\affiliation{MIT Kavli Institute and Department of Physics, 77 Massachusetts Avenue, Cambridge, MA 02139}

\author{George Ricker}
\affiliation{MIT Kavli Institute and Department of Physics, 77 Massachusetts Avenue, Cambridge, MA 02139}

\begin{abstract} % 148 words
We present two new rapidly oscillating Ap (roAp) stars, TIC\,198781841 and TIC\,229960986, discovered in TESS photometric data. The periodogram of TIC\,198781841 has a large peak at 166.506~d$^{-1}$ (1.93~mHz), with two nearby peaks at 163.412~d$^{-1}$ (1.89~mHz) and 169.600 d$^{-1}$ (1.96~mHz). These correspond to three independent high-overtone pressure modes, with alternating even and odd $\ell$ values. TIC\,229960986 has a high-frequency triplet centered at 191.641~d$^{-1}$ (2.218~mHz), with sidebands at 191.164~d$^{-1}$ (2.213~mHz) and 192.119~d$^{-1}$ (2.224 mHz). This pulsation appears to be a rotationally split dipole mode, with sideband amplitudes significantly larger than that of the central peak; hence, both pulsation poles are seen over the rotation cycle. Our photometric identification of two new roAp stars underscores the remarkable ability of TESS to identify high-frequency pulsators without spectroscopic observations.
\end{abstract}

% 1134 words in text + caption
\section{Introduction}

The Transiting Exoplanet Survey Satellite (TESS; \citealt{2015JATIS...1a4003R}) has revolutionized the field of asteroseismology with its short-cadence observations of over 200\,000 stars during its Prime Mission, from 2018 July to 2020 July \citep{2021ApJS..254...39G}. Of particular note are the rapidly oscillating chemically peculiar A (roAp) stars, whose high-frequency pulsations are easily identified using a continuously-observing, short-cadence space probe like TESS or Kepler that can produce a light curve with a high signal-to-noise ratio.

First discovered through targeted observations by \citet{1982MNRAS.200..807K}, roAp stars are a subclass of the Ap stars, which are strongly magnetic and exhibit enhanced abundances of rare earth metals \citep{doi:10.1146/annurev.aa.12.090174.001353}. roAp stars additionally exhibit high-overtone p~mode pulsations, with periods in the range $4.7 - 25.8$\,min ($\nu = 55 - 310$\,d$^{-1}$). Tens of new roAp stars have been identified from recent large-scale space- and ground-based photometric surveys (see \citealt{2021FrASS...8...31H} and references therein).

Much remains unexplained about these stars, however. Certain roAp stars pulsate at frequencies higher than those suggested by models (see, e.g., \citealt{2013MNRAS.436.1639C}), perhaps due to their pulsations being excited by a different mechanism than the typical $\kappa$~mechanism. Moreover, there have been roAp stars discovered that are cooler than the theoretical boundary of the instability strip, along with a pronounced dearth of stars toward the blue (hotter) edge of the instability strip \citep{2019MNRAS.487.3523C}. By finding more roAp stars, large surveys such as TESS can provide the statistics to determine the underlying cause of this discrepancy -- perhaps either from an observational bias or lower amplitudes in the pulsations of hotter stars.

\section{Observations and Data}

TIC\,198781841 was observed at 2-min cadence during TESS Sector~40, from 2021 June 24 to 2021 July 23. It was also observed during Sector~14 at 30-min cadence in the full-frame images (FFIs); however, the Nyquist limit of this data inhibits us from studying the high frequencies at which roAp stars normally pulsate. The 2-min cadence light curve for Sector 40 was generated by the Science Processing Operations Center pipeline at NASA Ames Research Center \citep{Jenkins-2015}. 

TIC\,229960986 was observed in 2-min cadence during TESS Sectors~40 and 41 (the latter from 2021 July 23 to 2021 August 20). It was also observed during much of Cycle 2, but at 30-min cadence. However, during Sectors 14 and 15, this target was present in the target pixel file of TIC\,229960976, which \textit{was} observed at 2-min cadence. The TESS Asteroseismic Science Operations Center (TASOC) used these data to construct a short-cadence light curve for TIC\,229960986. However, this light curve proved difficult to normalize, so we used only the data from Sectors 40 and 41 for our analysis.

These two stars were selected for manual review by a custom pipeline searching for high-amplitude peaks in the periodograms of all TESS short-cadence targets in a given sector. For TIC\,198781841, the roAp pulsations were flagged. For TIC\,229960986, the rotational frequency was flagged; human review uncovered a triplet of higher-frequency pulsations. The light curves and periodograms of the two stars are shown in Figure~\ref{fig:roAp_fft}. As these stars are hot enough to reside within the instability strip, we concluded that these are genuine roAp stars and characterized their pulsation modes.

\section{Mode Identification}
First, we identify the asymptotic (large) p~mode frequency separation, as in \citet{1995A&A...293...87K}:
\begin{equation}
    \Delta \nu_0 = \Delta \nu_{0,\odot} \sqrt{\frac{\overline\rho}{\bar{\rho}_\odot}}
\end{equation}
Here, $\Delta \nu_{0, \odot} = 135$~$\mu$Hz, and $\bar{\rho}_\odot = 1.41$~g\,cm$^{-3}$. The mean stellar density $\bar{\rho}$ can be obtained from calculating the luminosity $L$ and then using the tables from \citet{2013ApJS..208....9P} to determine the mass and radius for a given $T_{\rm eff}$ (obtained from the TESS Input Catalog; \citealt{2019AJ....158..138S}). To calculate $L$, we use the parallax and magnitude provided in the Gaia Early Data Release 3 \citep{2016A&A...595A...1G, 2021A&A...649A...1G}.

\subsection{TIC\,198781841}

This star has a Gaia magnitude $m_G$ of 10.75 and a parallax of $2.21\pm0.01$~mas; the latter yields a distance of $452\pm2$~pc. Thus, the absolute magnitude is $M_G = 2.474$. Neglecting a small bolometric correction, we assume $M_G \simeq M_{\rm bol}$, which yields $L \approx 8.1\,$L$_\odot$. For $T_{\rm eff} = 7725 \pm 250$~K, the table by \citet{2013ApJS..208....9P} suggests this is an A7/8V star, with a mass of $\sim$1.8~M$_\odot$ and a radius of $\sim$1.75~R$_\odot$, yielding $\bar{\rho} = 0.47$~g\,cm$^{-3}$. Thus, the asymptotic frequency separation $\Delta\nu_0 \approx 6.76$~d$^{-1}$ (78.2~$\mu$Hz). The observed separation of the three peaks in the second panel of Figure \ref{fig:roAp_fft} is 3.09 d$^{-1}$ (35.8~$\mu$Hz) -- nearly half the calculated $\Delta\nu_0$.

Unlike many recently-identified roAp stars (see, e.g., \citealt{2019MNRAS.487.3523C}), there exists no clear rotational signature within the span of the TESS data; the periodogram at low frequencies is consistent with noise. This indicates that this star could be a super-slowly rotating Ap (ssrAp) star \citep{2020A&A...639A..31M}, with a rotation period longer than the time span of the TESS data set. Such non-rotationally split triplets, representing high-overtone pressure modes, often alternate between even and odd spherical degree $\ell$. Because high $\ell$ modes tend to be strongly geometrically cancelled, the observed frequencies arise from alternating even and odd low $\ell$ values (e.g., $\{2,1,2\}$ or $\{0,1,0\}$). These are high-overtone modes ($n \gtrsim 20$, based on the large frequency separation; see, e.g., the Introduction of \citealt{2005MNRAS.356..671O}).

\subsection{TIC\,229960986}

This star has $m_G = 10.50$ and a parallax of $2.27\pm0.04$~mas, corresponding to a distance of $441\pm8$~pc; this yields $M_G = 2.28$. Assuming again that $M_G \simeq M_{\rm bol}$, we obtain $L \approx 9.64$\,L$_\odot$. For TIC~$T_{\rm eff} = 8150 \pm 145$~K, the table of \citet{2013ApJS..208....9P} suggests this is closest to an A5V star, with a mass of $\sim$1.9~M$_\odot$ and a radius of $\sim$2~R$_\odot$, yielding $\bar{\rho} = 0.34$~g\,cm$^{-3}$. The asymptotic frequency separation $\Delta\nu_0 \approx 5.69$~d$^{-1}$ (65.8\,$\mu$Hz).

\begin{figure*}
    \centering
    \includegraphics[width=.95\linewidth]{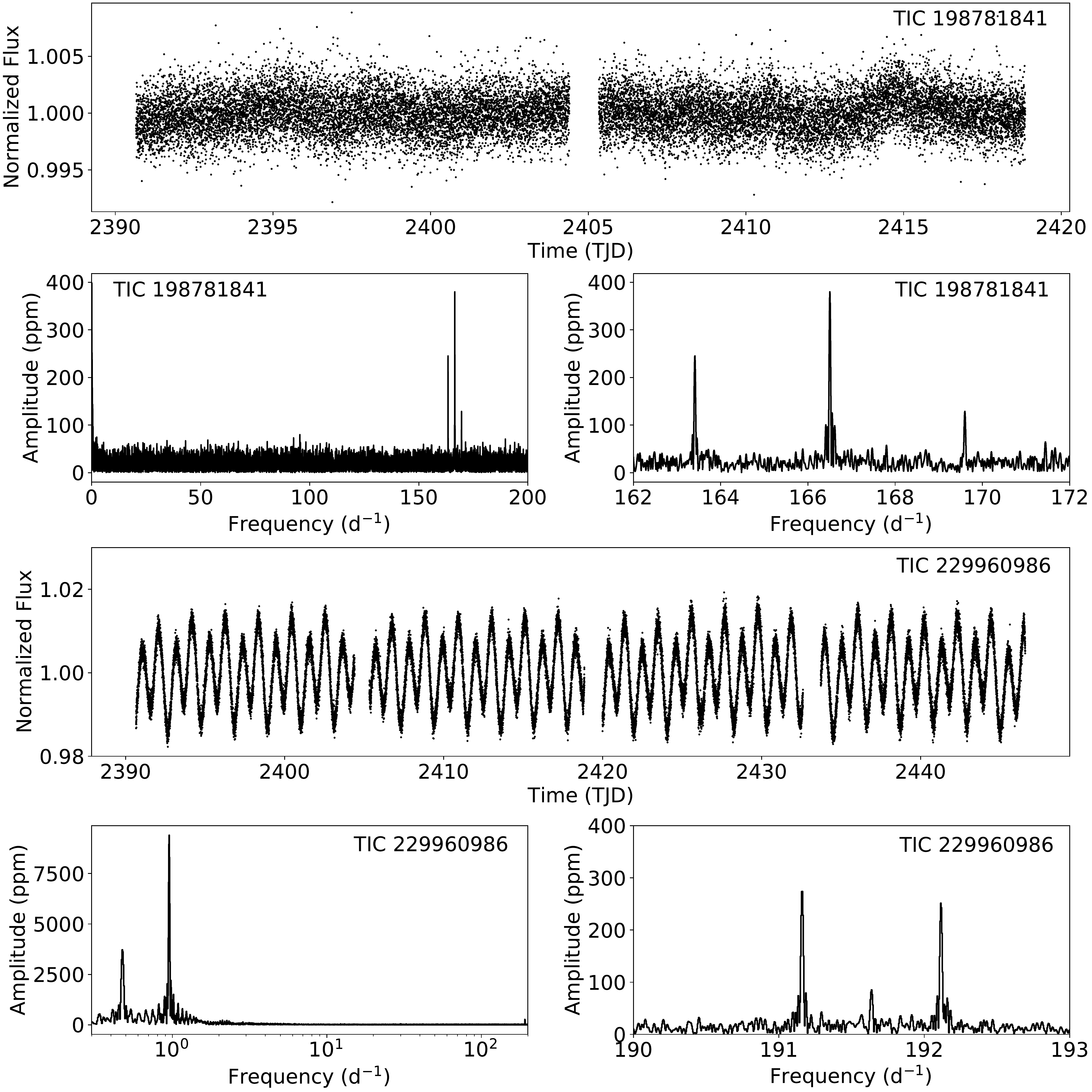}
    \caption{Light curves and periodograms for the two new roAp stars. \textit{Row 1}: The 2-min cadence light curve from Sector~40 for TIC\,198781841. \textit{Row 2}: This light curve's periodogram from a Discrete Fourier Transform (DFT; see, e.g., \citealt{1985MNRAS.213..773K}), with a zoom into the roAp pulsations. \textit{Row 3}: The 2-min cadence light curve from Sectors 40 and 41 for TIC\,229960986. \textit{Row 4}: The periodogram of this light curve; the left panel is plotted on a semi-logarithmic scale to highlight the rotational frequency; the right panel shows the roAp pulsations.}
    \label{fig:roAp_fft}
\end{figure*}

Unlike TIC\,198781841, this star \textit{does} have a clear rotational period of $P_{\rm rot} = 2.0946\pm0.0002$~d, shown in the bottom panel of Figure \ref{fig:roAp_fft}. The equally-spaced high-frequency triplet is split by the rotational frequency $0.4774$~d$^{-1}$ to within 1$\sigma$. This is a signature of an oblique dipole mode with $\ell$=1, as we expect $2\ell+1$ peaks for a given dipole $\ell$ mode, unless the $|m|=2$ components are within the noise. Since $P_{\rm rot} \lesssim 0.1\Delta\nu_0$, the three frequencies cannot be alternating even and odd degree modes, as was the case for the previous star.

\section{Conclusion}

We have used TESS photometry to discover and characterize two new roAp stars. Large sky surveys such as TESS have considerably reduced the difficulty of detecting such stars. Future work will focus on obtaining spectra to identify chemical overabundances in these two stars and searching the TESS data for new pulsators.

\facility{TESS}
%%references WC: 227
\bibliography{roAp}{}
\bibliographystyle{aasjournal}
\end{document}